\documentclass[aps,prl,twocolumn,amsfonts,amsmath,floatfix,showpacs,superscriptaddress,footnoteintext]{revtex4}

\usepackage[final]{graphicx}
\usepackage{color}

\newcommand{\be}{\begin{equation}} \newcommand{\ee}{\end{equation}}

\begin{document}

\title{Exact results for classical Casimir interactions:\\ Dirichlet and Drude
  model in the sphere-sphere and sphere-plane geometry}

\date{\today}

\author{Giuseppe Bimonte}
\affiliation{Dipartimento di Scienze
Fisiche, Universit{\`a} di Napoli Federico II, Complesso Universitario
MSA, Via Cintia, I-80126 Napoli, Italy}
\affiliation{INFN Sezione di
Napoli, I-80126 Napoli, Italy }

\author{Thorsten Emig}
\affiliation{Laboratoire de Physique
Th\'eorique et Mod\`eles Statistiques, CNRS UMR 8626, B\^at.~100,
Universit\'e Paris-Sud, 91405 Orsay cedex, France}

\begin{abstract}
  Analytic expressions that describe Casimir interactions over the
  entire range of separations have been limited to planar
  surfaces. Here we derive analytic expressions for the classical or
  high-temperature limit of Casimir interactions between two spheres
  (interior and exterior configurations), including the sphere-plane
  geometry as a special case, using bispherical coordinates. We
  consider both Dirichlet boundary conditions and metallic boundary
  conditions described by the Drude model. At short distances,
  closed-form expansions are derived from the exact result, displaying
  an intricate structure of deviations from the commonly employed
  proximity force approximation.
\end{abstract}


\maketitle

The collective action of fluctuation ``van der Waals'' forces between
individual atoms leads to forces between macroscopic surfaces
\cite{Parsegian,bordag}. The fluctuations can be of thermal
\cite{deGennesFisher} and/or quantum \cite{Casimir48} nature.  An
important characteristic of these forces is their non-additivity which
complicates the study of Casimir interactions in the thermodynamic
limit involving a macroscopic number of particles.  An important
exception is the interaction of {\it planar, parallel} surfaces where
symmetry allows for an exact solution, the so-called Lifshitz formula
\cite{Lifshitz}, which is a landmark in the physics of fluctuation
forces. More complicated shapes have been studied at sufficiently
short surface separations by the proximity force approximation (PFA)
that is based on the Lifshitz formula, treating curvature by summing
over planar surface elements \cite{Derjaguin}. The thermal Casimir interaction
between spherical particles in a critical fluid has been computed using conformal
invariance for nearly touching and widely separated particles \cite{Eisenriegler}.
More recently, a number
of analytical and numerical approaches have been developed to deal
with the non-additivity, most notably implementations of concepts from
scattering theory \cite{universal,Lambrecht_SLP,Johnson_SLP}. However,
it remains unclear to what extend and precision these approaches can
handle the practically important limit of short distances. Attempts to
obtain analytic predictions beyond the PFA have been limited to
compute first order corrections from a gradient expansion
\cite{fosco,Bimonte2011,Bimonte2012}.

The difficulty to compute deviations from the simple PFA can be traced
back to the long-range nature of fluctuation forces which is poorly
treated by the PFA. With increasing spatial dimensions, fluctuations
decay more strongly, and the PFA is expected to become more reliable.
The situation resembles the situation in statistical mechanics when
phase transitions are studied by mean field theory, treating
fluctuations poorly. In this context, the exact solution of the
two-dimensional Ising model \cite{Onsager} helped understanding phase
transitions and inspired several developments in the theory of
critical phenomena and related fields.  This demonstrates the
importance of exact solutions for a better understanding of complex
phenomena. In particular, several forms for non-additivity induced
deviations from the PFA have been hypothesized before, depending on
boundary conditions and temperature
\cite{Emig2008,Lambrecht_SLP,reynaud}. Here we derive an exact
solution for the interaction between spheres of different radii that
displays a much richer structure for the Casimir interaction in the
high temperature limit than previously assumed.

In this Letter, we obtain analytic expressions for the
high-temperature limit of the Casimir interactions between two spheres
(interior and exterior configurations), including the sphere-plane
geometry as a special case.  We consider both a scalar field with
Dirichlet boundary conditions and metallic boundary conditions
described by the Drude model. The key tool of our approach is a
bispherical coordinate system \cite{MF}, which allows for a simple
solution of the scattering problem of two spheres in the static limit
(Laplace equation). The high-temperature sphere-plate problem has been
analyzed before in the large distane limit \cite{emig} and by
large-scale numerics at short distances, including up to $5000$
partial wave orders \cite{reynaud}. Our exact solution is universal,
i.e., material independent and displays full agreement with these two
limiting cases. We derive from the exact result a closed-form
short-distance expansions that reveals an intricate structure of
deviations from the PFA.  It turns out that for Dirichlet boundary
conditions, the classical Casimir {\it force} can be expanded as a
Laurent series for small surface-to-surface separations $L$. For Drude
metallic boundary conditions, the structure of the corrections to PFA
is more complicated than suspected before \cite{reynaud}, as it
involves a double logarithm of $L$, as well as powers of $\ln L$
multiplied by powers of $L$.  In both cases the leading correction to
the PFA energy is represented by a term proportional to $\ln L$, with
a common coefficient that had been computed earlier \cite{Bimonte2012}
by using a recently proposed gradient expansion of the Casimir energy
\cite{fosco,Bimonte2011}.  However, for experimentally accessible
separations, the interaction of Drude metals is dominated entirely by
the double logarithmic term.

\begin{figure}[h]
\includegraphics[width= \columnwidth]{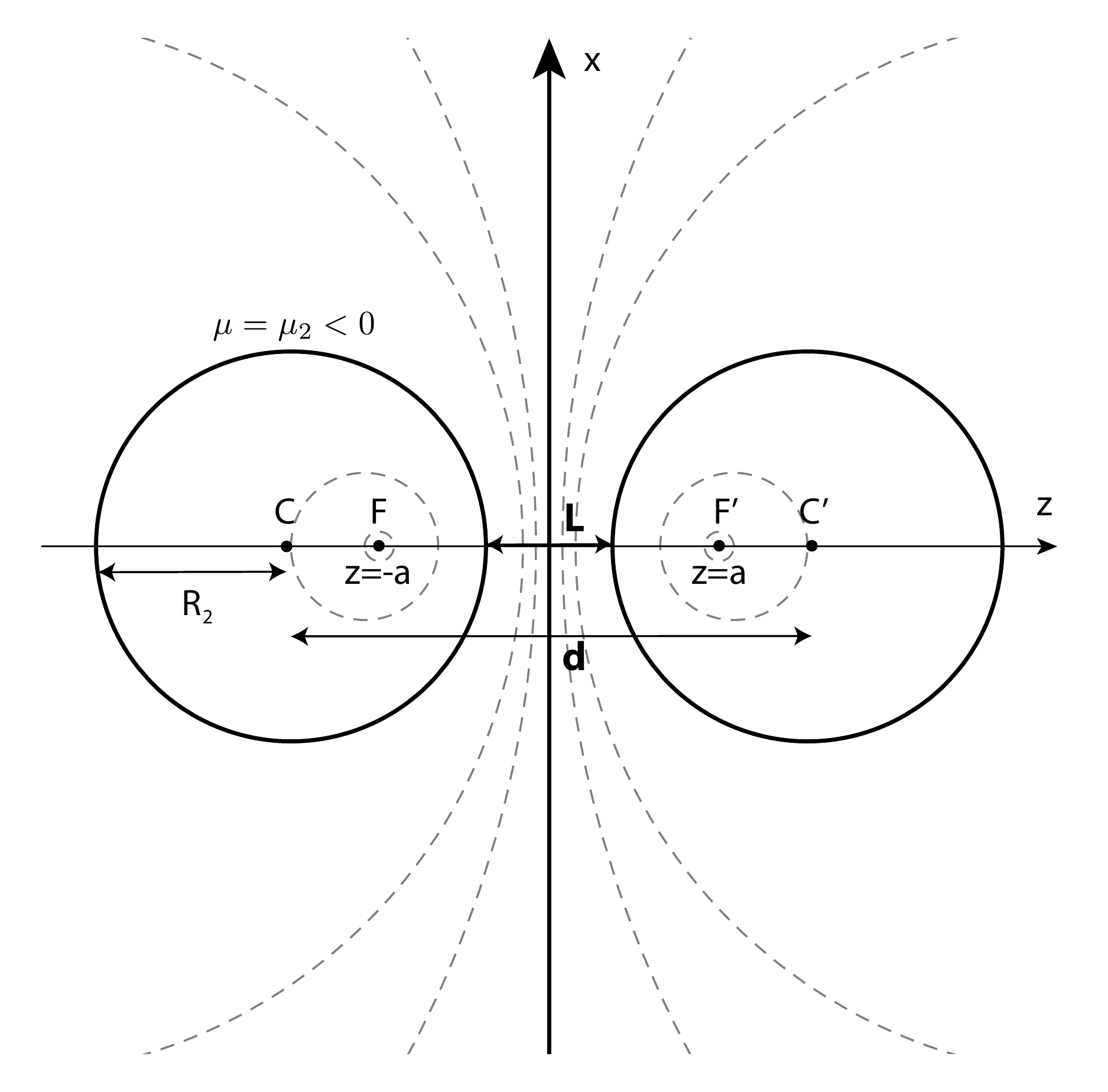}
\caption{Geometry of two spheres and sphere-plate. Shown are the
  centers ($C$, $C'$) of the spheres and their foci ($F$, $F'$). The
  dashed curves correspond to curves of constant bispherical
  coordinate $\mu$ with $\mu=\mp \pi,\, \mp \pi/2,  \,
  \mp \pi/10, \, \mp \pi/20$ with the upper (lower) sign for $z<0\,
  (z>0)$, beginning with the smallest sphere.}
\label{fig:geometry}
\end{figure}

{\it Geometry, coordinates and eigenfunctions} --- We consider two
spheres of radii $R_1$ and $R_2$ with surface-to-surface distance
$L$. This geometry is conveniently parametrized in bispherical
coordinates $(\mu,\eta,\varphi)$ \cite{MF}, defined by
$(x,y,z)=a(\sin\eta \cos\varphi,\sin\eta\sin\varphi,\sinh
\mu)/(\cosh\mu-\cos\eta)$ where $z=\pm a$ are the foci of the two
spheres defined by $\mu=\mu_1>0$ and $\mu=\mu_2<0$, respectively, see
Fig.~\ref{fig:geometry}.  The spheres have radii $R_1=a/\sinh\mu_1$
and $R_2=-a/\sinh\mu_2$ and distances $L_1=a\coth\mu_1$ and
$L_2=-a\coth\mu_2$ from the origin. The center-to-center distance is
$d=L_1+L_2$. It is useful to express the $\mu_\alpha$ in terms of the
natural geometrical scales, $\mu_1=\text{arccosh} \lambda_1$,
$\mu_2=-\text{arccosh} \lambda_2$ with
$\lambda_\alpha=[L^2+2(L+R_\alpha)(R_1+R_2)]/[2R_\alpha(L+R_1+R_2)]$.
We are interested in the classical Casimir interaction which is
determined by the scattering of the spheres at zero frequency which is
determined by the Laplace equation. The latter is separable in
bispherical coordinates and its Green's function can be expanded as
$G_0(\mu,\eta,\varphi;\mu',\eta',\varphi') = \sum_{l=0}^\infty
\sum_{m=-l}^l \phi_{lm}^\text{reg} (\mu_<,\eta,\varphi)
\phi_{lm}^\text{out*} (\mu_>,\eta',\varphi')$ where $\mu_{<(>)}$ is
the smaller (larger) of $\mu$ and $\mu'$ and the regular and outgoing
eigenfunctions are
  \begin{equation}
    \label{eq:eigenfunctions}
     \phi_{lm}^\text{reg/out} =
  \sqrt{\frac{\cosh\mu-\cos\eta}{a(2l+1)}} Y_{lm}(\eta,\varphi)
  e^{\pm(l+1/2)\mu} \,
  \end{equation}
  for $l\ge 0$, $m=-l,\ldots,l$. Scattering solutions for various
  boundary conditions can be expanded in these eigenfunctions, leading
  to the zero-frequency (static) scattering amplitude
  $T^{(\alpha)}_{lml'm'}$  of sphere $\alpha$ by which we denote the matrix
  elements of the operator $\hat T^{(\alpha)}$ in the bispherical
  basis. The Casimir energy is expressed in the scattering approach
  \cite{universal} in terms of the scattering amplitudes and
  translation operators that translate the scattering solution from
  the coordinate of one object to the one of the other object. Here,
  for both spheres the scattering amplitude is expressed in the same
  bispherical coordinate system and hence the translation operators
  become the identity operators, yielding for the classical Casimir
  energy (or $0^\text{th}$ order Matsubara term),
\begin{equation}
  \label{eq:energy_general}
  E = \frac{k_B T}{2} \ln \det [1 -\hat N]  \quad \text{with} \quad
  \hat N=
  \hat T^{(1)} \hat T^{(2)} \, .
\end{equation}
It is important to notice that the energy depends only on the
equivalence class $[[N]]$ formed by all matrices that represent the
operator $\hat N$, i.e., two elements $N$ and $M N M^{-1}$ of the
class differ by an (invertible) similarity transformation $M$. In the
following we denote by $[[\ldots ]]$ the equivalence class of the
matrix enclosed by the brackets.

{\it Dirichlet boundary conditions} ---
The scattering amplitudes for Dirichlet boundary conditions
$\phi(\mu=\mu_\alpha)=0$ on the
spheres follows from Eq.~(\ref{eq:eigenfunctions}) \cite{universal}.
They assume the simple diagonal and $m$-independent form
 $T_{lml'm'}^{(\alpha)}=-\exp(\mp(2l+1)\mu_\alpha) \delta_{ll'} \delta_{mm'}$ with $-$ for $\alpha=1$
and $+$ for $\alpha=2$. Hence, the determinant in
Eq.~(\ref{eq:energy_general}) can be easily computed and we get for the
classical Casimir energy of two Dirichlet spheres the exact result
\begin{equation}
  \label{eq:D-case_energy}
  E^\text{(D)} = \frac{k_B T}{2} \sum_{l=0}^\infty (2l+1) \ln \left[ 1 -
  Z^{2l+1} \right]
\end{equation}
which depends only via the single variable
$Z=[\lambda_1+\sqrt{\lambda_1^2-1}]^{-1}
[\lambda_2+\sqrt{\lambda_2^2-1}]^{-1}$ on the geometrical scales. For
$|Z|<1$, including the range of physically meaningful
parameters, the energy $E^\text{(D)}$ is an analytic function
\footnote{This can be seen by expanding the logarithm and rewriting
  the energy as a power series in $Z$ whose coefficients $a_\nu$ do
  not increase faster than linearly with $\nu$.}. Large distances $L$
correspond to small $Z$ with $Z=R_1R_2/L^2+{\cal O}(L^{-3})$ and small
distances to $Z$ close to unity with $Z=1-\sqrt{2L/R_1+2L/R_2}+{\cal
  O}(L)$.

We note that the configuration of two spheres with one ($R_1$) inside
the other ($R_2$) \cite{Zaheer_2010a,Zaheer_2010b} can be computed
exactly by the same method. The classical energy is given again by
Eq.~(\ref{eq:D-case_energy}) but now with
$Z=[\lambda_2+\sqrt{\lambda_2^2-1}]/ [\lambda_1+\sqrt{\lambda_1^2-1}]$
with $\lambda_1= [-L^2+2(L+R_1)(-R_1+R_2)]/[2R_1(-L-R_1+R_2)] >
\lambda_2= [L^2+2(-L+R_2)(-R_1+R_2)][2R_2(-L-R_1+R_2)]$ where $L$ is
the closest surface-to-surface separation of the spheres so that their
centers have a distance $d=R_2-R_1-L>0$. For two concentric spheres one
has $Z=R_1/R_2$.

{\it Drude model} --- In general, a Drude metal has to be described
within electromagnetic scattering theory. However, in the classical
limit, it can be shown that only the electric or TM modes contribute
to the Casimir energy \cite{emig}. We shall use this observation below
to map the interaction of Drude metals to a scalar field problem that
is similar to the one with Dirichlet conditions. To simplify
notations, we focus here on the sphere-plate geometry (with the sphere
described by $\mu=\mu_1>0$).  When the
operator $\hat N$ that yields the classical energy for the Drude model
is expressed in a {\it spherical} wave basis with $C$ as the origin (see
Fig.~\ref{fig:geometry}), it has the form \cite{emig}
\begin{equation}
  \label{eq:N_sphere-plane_spherical}
  \hat N = \left[\left[ \frac{(l+l')!}{(l+m)!(l'-m)!}
      \left(Z+Z^{-1}\right)^{-l-l'-1} \delta_{mm'}
\right]\right]
\end{equation}
where we used that $R/(L+R)=2/(Z+1/Z)$ for a plate and a sphere of
radius $R$. The same matrix is obtained for a scalar field with
Dirichlet conditions but with one important difference: While in the
Dirichlet case monopoles with $l=0$ are included, in the Drude case
monopoles are excluded since there are no corresponding spherical
vector waves.  Physically, this feature is a consequence of the fact
that an {\it isolated} compact object has zero total charge, i.e., no
monopole. The next important observation is that a given spherical
multipole of order $(l,m)$ decomposes into infinitely many bispherical
multipoles $(l',m)$ with $m$ conserved and hence in the bispherical
basis the difference between Dirichlet and Drude conditions is more
complex. Since monopoles $l=0$ occur only in the sector for $m=0$,
it is sufficient to reconsider the part of the energy coming from
$m=0$ modes, and to transfer the Dirichlet energy unchanged for $m\neq
0$, giving the classical Casimir energy for the Drude case
\begin{equation}
  \label{eq:split}
  E^\text{(Dr)}=
E^\text{(Dr)}_{m=0}+\frac{k_B T}{2}\sum_{l=1}^{\infty} 2l
\ln(1-Z^{2l+1}) \, ,
\end{equation}
where $E^\text{(Dr)}_{m=0}$ is given by Eq.~(\ref{eq:energy_general})
with $\hat N$ from Eq.~(\ref{eq:N_sphere-plane_spherical}) with $m=0$
and restricted to $l,l'\ge 1$, the equivalence class of which we
denote by $\hat N_0$ in the following. To compute the determinant in
Eq.~(\ref{eq:energy_general}), we perform two transformations on the
matrix that represents $\hat N_0$: (i) A translation from the {\it
  spherical} wave basis $S_{C}$ with origin $C$ to a {\it
  spherical} wave basis $S_{F}$ with origin $F$, see Fig.~\ref{fig:geometry}. (ii) A
conversion from the basis $S_{F}$ to the {\it bispherical} wave basis
$BS_{F}$ with the same origin $F$. The translation (i) over the
distance $Z R$ corresponds to a similarity transformation with a matrix
which has non-zero elements only for $l \ge l' \ge 1$, given by
\begin{equation}
\label{eq:matrix_V}
{\cal V}^{S_{C}\to S_{F}}_{ll'}(Z R)=
\frac{ (Z\,R) ^{\,l-l'}}{(l-l')!} \frac{l!}{l'!}\sqrt{\frac{2l'+1
  }{2l+1 }} \, . \nonumber
\end{equation}
The inverse of this matrix is obtained by $ZR \to -ZR$.
After this transformation, we can write   $\hat
N_0$ as
\begin{equation}
  \label{eq:N0_after_i}
  \hat N_0 = \left[\left[
\begin{array}{ll}
-(-1)^l Z^{l+l'+1} & \text {if}\quad l>l' \\
\left( \frac{l'!}{l!(l'-l)!} - (-1)^l \right) Z^{l+l'+1} & \text {if}
\quad l\le l'
\end{array}
  \right]\right]
\end{equation}
The conversion (ii) corresponds to a similarity transformation with a
matrix which has non-zero elements only for $l'\ge l\ge 1$, given by
\begin{equation}
{\cal V}^{S_{F}\to BS_{F}}_{ll'}=  \frac{ (-1)^l}{\sqrt{2l+1}} \left[R
  \left(\frac{1}{Z}-Z\right)\right]^{l+1/2}\!\!\!\frac{l'!}{l!
  (l'-l)!} \, . \nonumber
\end{equation}
With this transformation, the equivalence class can be expressed as
\begin{equation}
  \label{eq:N0_after_ii}
  \hat N_0 = \left[\left[ Z^{2 l'+1}\left(\delta_{ll'}+ (1-Z^2)(1-Z^{2l'})  \right)
  \right]\right]
\end{equation}
The latter expression allows for a direct computation of $\det (1-\hat
N_0)$. The first part in Eq.~(\ref{eq:N0_after_ii}) is diagonal and
can be easily factorized so that it yields a contribution $(k_B
T/2)\sum_{l=1}^\infty \ln(1-Z^{2l+1})$ to the energy $E^\text{(Dr)}_{m=0}$.
The second part in Eq.~(\ref{eq:N0_after_ii}) depends only the column
index $l'$ and hence is a matrix with equal rows. For a matrix $A$ of
this type one has $\det (1-A)=1- \text{tr} A$. Hence the contribution
of the second part  in Eq.~(\ref{eq:N0_after_ii}) to the energy is
given by the trace of this part divided by $1-Z^{2l+1}$ due to
the factorization of the first part of $1-\hat N_0$. Combining the two
parts from Eq.~(\ref{eq:N0_after_ii}) with Eq.~(\ref{eq:split}) we
obtain the following {\it exact} expression for the
classical Casimir energy of a Drude sphere and plate
\begin{widetext}
\begin{equation}
\label{eq:E_Drude}
E^\text{(Dr)} = \frac{k_B T}{2} \left[ \sum_{l=1}^{\infty} (2l+1)
  \ln(1-Z^{2l+1})+\ln\left(1-(1-Z^2)\sum_{l=1}^{\infty}Z^{2l+1}\frac{1-Z^{2l}}{1-Z^{2l+1}}\right)\right] \;.
\end{equation}
\end{widetext}
Applying similar arguments as in the Dirichlet case, it can again be
shown that $E^\text{(Dr)}$ is an analytic function for $|Z|<1$. We note
that the effect of eliminating monopole fluctuations from the energy
for Dirichlet conditions is not only a sum starting at $l=1$ in
$E^\text{(D)}$ but the occurrence of a second logarithmic term. This
is a consequence of the coupling of monopoles to higher order multipoles.

\begin{figure}[h]
\includegraphics[width= \columnwidth]{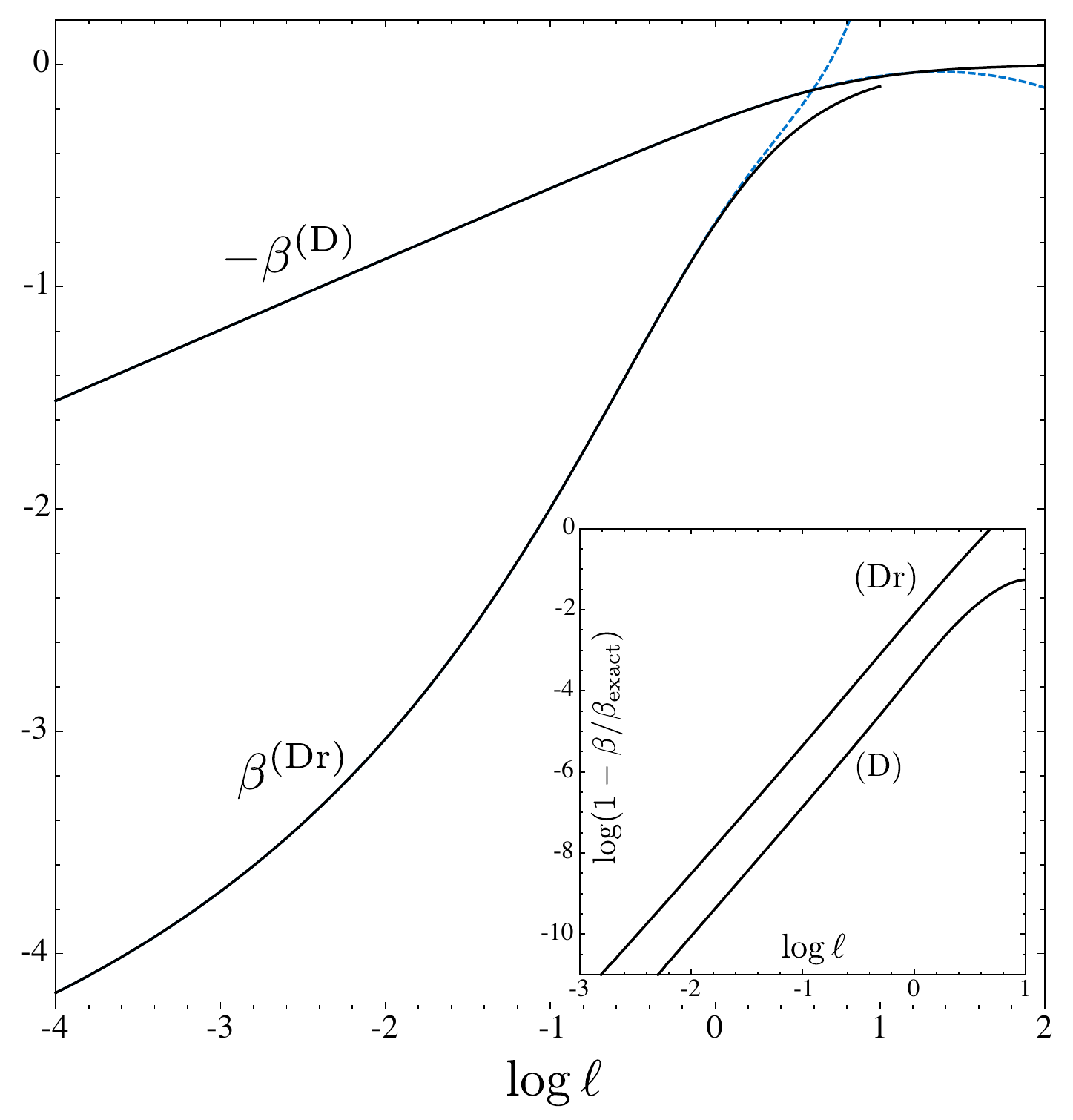}
\caption{Deviations from proximity force approximation in the
  sphere-plate geometry, measured by the function $\beta(\ell)$ defined
  by $E^\text{(D)/(Dr)}=-\zeta(3) k_B T/8 [1/\ell +
  \beta^\text{(D)/(Dr)}(\ell)]$ with $\ell=L/R$. Shown are the exact
  results for Dirichlet conditions [Eq.~\eqref{eq:D-case_energy}] and
  the Drude model [Eq.~\eqref{eq:E_Drude}] as solid lines and the
  short distance expansions of
  Eqs.~\eqref{eq:D_energy_small_distance},
  \eqref{eq:Drude_energy_small_distance} as dashed lines as function
  of the logarithm (with base $10$) of $\ell$. Note that
  $\beta^\text{(D)}>0$ and $\beta^\text{(Dr)}<0$ in the shown
  range. The inset shows the logarithmic relative difference between
  the exact result $\beta_\text{exact}$ from
  Eqs. \eqref{eq:D-case_energy}, \eqref{eq:E_Drude} and the $\beta$
  obtained from the short distance expansions of
  Eqs.~\eqref{eq:D_energy_small_distance},
  \eqref{eq:Drude_energy_small_distance}.}
\label{fig:beta}
\end{figure}

{\it Expressions at short distances} --- With exact expressions for
the Casimir energies in Dirichlet and Drude model available, one can
compute explicitly the interaction in the experimentally important
limit of short distances $L\lesssim R$.  Since this limit corresponds
to $Z$ close to unity, we compute the series in
Eqs.~(\ref{eq:D-case_energy}), (\ref{eq:E_Drude}) using the Abel-Plana
formula \cite{bordag}. We set $Z=\exp(-\mu)$ and expand for small $\mu$ where
$\mu=\ln(1+\ell+\sqrt{\ell(2+\ell)})$, $\ell=L/R$, in the
sphere-plate geometry. For Dirichlet conditions, we obtain
\begin{widetext}
\begin{equation}
  \label{eq:D_energy_small_distance}
  E^\text{(D)} = \frac{k_B T}{2} \left[   -\frac{\zeta(3)}{2}
    \frac{1}{\mu^2}
+\frac{1}{12} \ln \mu +
   \frac{1}{8} - \gamma_0 -\frac{7}{2880} \mu^2
   -\frac{31}{725760} \mu^4 + {\cal O}(\mu^6)\right]
\end{equation}
with the constant $\gamma_0=0.174897$ that is given by an integral
\footnote{$\gamma_0=-i\int_0^\infty dt
  [(1+2it)\ln(1+2it)-\text{c.c.}]/e^{2\pi t-1}$}, and for the Drude
model we get
\begin{equation}
  \label{eq:Drude_energy_small_distance}
  E^\text{(Dr)} =  E^\text{(D)} +\frac{k_B T}{2} \left[
    \ln(\gamma_1-\ln \mu) +
\frac{1}{6} \,\frac{-\gamma_2+\ln \mu}{-\gamma_1+\ln \mu} \mu^2
- \frac{1}{180} \, \frac{\gamma_3-\gamma_4\ln \mu +\ln^2 \mu}
{(-\gamma_1+\ln \mu)^2} \mu^4 + {\cal O}(\mu^6)
\right] \, ,
\end{equation}
\end{widetext}
where the constants $\gamma_1=1.270362$, $\gamma_2=1.35369$,
$\gamma_3=1.59409$,
$\gamma_4=2.51153$ are given by integrals that can
be computed easily numerically. We used $\mu$ as the variable for the
expansion since it provides a very accurate result at even larger
distances, see Fig.~\ref{fig:beta}. A general feature of both the Dirichlet and
Drude energy is their dependence on only $\ln \mu$ and {\it even}
powers of $\mu$. This shows that the energies depend only on $\ln
\ell$ and {\it integer} powers of $\ell$.  When the {\it force} for
the Dirichlet case is expanded in $\ell$, it becomes a Laurent series
starting with $1/\ell^2$, i.e., there are no logarithmic terms for the
force. For the Drude model, there are logarithmic terms in the force,
and the most convenient form to express the short distance expansion
appears to be the one in
Eq.~(\ref{eq:Drude_energy_small_distance}). The leading correction to
the PFA is the same term $\sim \ln\mu$ in both models. However, at
realistic distances, the double logarithmic term in
Eq.~(\ref{eq:Drude_energy_small_distance}) dominates and therefore the
two models show rather distinct behavior, see Fig.~\ref{fig:beta}.

It is instructive to discuss the different behavior of the classical
Casimir energies for Dirichlet and Drude boundary conditions.  For
large separations, the interactions display different asymptotic
behaviors for the two boundary conditions.  In the experimentally most
relevant sphere-plate geometry Eq.~\eqref{eq:D-case_energy} predicts
the known $1/\ell$ fall-off rate for the Dirichlet energy, while in
the Drude case Eq.~\eqref{eq:E_Drude} gives the characteristic
$1/\ell^3$ decay \footnote{The same $1/\ell^3$ fall-off rate, but with
  twice the numerical coefficient, applies to the plasma model in the
  limit where the plasma wavelength $\lambda_p\ll R$ \cite{emig}.}.
The slower decay rate in the Dirichlet case results from monopole
contributions that are absent in the Drude case.  While a Dirichlet
scalar field is in general not expected to describe the Casimir
interaction between metals, in the high-temperature limit the
difference between the two universal interactions for Dirichlet and
Drude conditions suggests an interesting physical interpretation.  In
fact, the high-temperature limit of the Casimir interaction provides
the dominant contribution to the full quantum Casimir interaction at
{\it finite} temperatures in the limit of sufficiently {\it large
  separations} \cite{Parsegian,Lifshitz}. To understand if the
Dirichlet ($1/\ell$) or Drude ($1/\ell^3$) decay describes experiments
with conductors at large distances, one must bear in mind that the
conductors used in Casimir force measurements are always connected
to a charge reservoir to compensate stray charges that might otherwise be present
on the surfaces.  In the static limit, the surface electric potential
of such a conductor is constant (Dirichlet boundary conditions) and
their total charge can fluctuate.  In contrast, Drude metallic boundary
conditions instead describe {\it ungrounded} charge-neutral
conductors. From this we conclude that the quantum Casimir force
between grounded conductors at finite temperatures should decay
according to the Dirichlet case at large distances since the energy is
than dominated by the lowest Matsubara mode, i.e., the classical
energy \cite{Parsegian,Lifshitz}. To discriminate between the
asymptotic $1/\ell$
and $1/\ell^3$ decay, it is necessary to consider sphere-plate
separations comparable to the sphere-radius or larger. Interestingly,
our short distance expansions in the classical limit suggest a distinct
behavior due to grounding at even shorter separations which, however,
is certainly modified due to quantum corrections.

We thank R. L. Jaffe and M. Kardar for valuable discussions.  This
research was supported by the ESF Research Network CASIMIR.

\end{document}